\documentclass[aps,preprint,pra]{revtex4}

\usepackage{graphicx}
\usepackage{amsmath}

\begin{document}

\title{Influence of substitution on the optical properties of functionalized
pentacene monomers and crystals: Experiment and theory}

\author{Y. Saeed$^1$, K. Zhao$^1$, N. Singh$^{1,2}$,
R. Li$^1$, J. E. Anthony$^3$, A. Amassian$^1$, and U. Schwingenschl\"ogl$^1$}

\affiliation{$^1$KAUST, Physical Science \& Engineering division, Thuwal
23955-6900, Kingdom of Saudi Arabia}

\affiliation{$^{2}$Solar and Photovoltaic Energy Research Center, KAUST, Thuwal 23955-6900, Kingdom of Saudi Arabia}

\affiliation{$^3$Department of Chemistry, University of Kentucky, Lexington,
Kentucky 40506-0055}

\begin{abstract}
The influence of solubilizing substitutional groups on the electronic
and optical properties of functionalized pentacene molecules and
crystals have been investigated. Density functional theory
is used to calculate the electronic and optical properties of pentacene,
TIBS-CF$_{3}$-pentacene, and TIPS-pentacene. The results are compared
with experimental absorption spectra of solutions and the complex
dielectric function of thin films in the 1 eV to 3 eV energy range. In all cases, the band gaps
of the isolated molecules are found to be smaller than those of the crystals. The
absorption spectra and dielectric function are interpreted in terms of the transitions
between the highest occupied molecular orbitals and lowest unoccupied molecular
orbitals. The bands associated to C and Si atoms connecting the functional side
group to the pentacene in the (6,13) positions are found to be the main contributors
to the optical transitions. The calculated dielectric functions of thin films agree with 
the experimental results. A redshift is observed in crystals as compared to molecules
in experiment and theory both, where the amplitude depends on the packing structure.
\end{abstract}

\maketitle

\section{Introduction}
The design of molecular semiconductors is increasingly important for the development of organic
electronics and organic photovoltaics (OPV) \cite{Zaumseil,WANG,DODABALAPOR}. Early on pentacene
had proven to be one of the best performing molecular semiconductors,
as vacuum-deposited organic thin film transistors have achieved a mobility as high
as 6 cm$^2$ V$^{-1}$ s$^{-1}$ \cite{Li,Jurchescu,Kim}.
However, it did not lend itself well to solution processing,
which is believed to be key for low-cost manufacturing of organic semiconductors. In recent years, chemical modification of the acene has made it possible to
overcome the low solubility and poor stability in solution, whilst
maintaining or enhancing the inter-molecular orbital overlap \cite{Anthony,Subramanian}. 

Functional substitution of pentacene has been shown to induce favorable
crystal packing motifs for both electronic and OPV applications  \cite{Subramanian,Shu}.
For example, pentacene without substitution shows a two-dimensional (2D) herringbone packing
motif (see Fig. 1a), while the popular compound (6,13)-bis(tri-iso-propyl-silyl-ethynyl)-pentacene
(TIPS-Pn) shows a brickwork 2D crystal packing (see Fig. 1c) \cite{Ostroverkhova}.
The latter is currently one of the organic semiconductors
exhibiting the highest field-effect mobility \cite{Jackson}, with recent
carrier mobility reports exceeding 4 cm$^2$ V$^{-1}$
s$^{-1}$ \cite{Bao-Nature}. When changing substitution from
tri-iso-propyl-silyl-ethynyl to tri-iso-butyl-silyl-ethynyl and introducing
a tri-fluoro-methyl group on the acene backbone to modify the energy
levels, to get 2-tri-fluoro-methyl-(6, 13)-bis-(tri-iso-butyl-silyl-ethynyl)-pentacene
(TIBS-CF$_{3}$-Pn), the crystal packing changes to one-dimensional (1D) sandwich
herringbone (see Fig. 1b). This molecule is found to perform as one of the best non-fullerene
acceptor molecules when mixed with P3HT donor polymer, yielding a power
conversion efficiency of 1.28\% \cite{Shu}.

The substitutional chemistry employed affects the electronic properties
of the monomer as well as  of the solid state material itself. Meng \emph{et al.} \cite{Meng} have demonstrated
that adjusting the alkyl substitution to the four terminal positions
(2, 3, 9, and 10) of the pentacene chromophore shifts the energies of both the highest
occupied molecular orbital (HOMO) and lowest unoccupied molecular
orbital (LUMO) without significantly changing the gap between these
two. When substituting all hydrogen atoms of pentacene with fluorine
atoms some interesting changes of the extinction coefficient are
found as the optical band gap is redshifted \cite{Hinderhofer}. Recently,
Lim \emph{et al.} \cite{lim} have shown that the HOMO--LUMO energy levels
can be tuned by varying the number of nitrile groups in cyano-pentacene
substitution. From the above reports, a close relationship
appears to exist between the substitution on the pentacene chromophore and
its electronic and optical properties. To tailor and improve these
properties, one should first understand the correlation
between the chemical modification (like silyl-ethynyl substitution as in the cases of
TIPS-Pn and TIBS-CF$_3$-Pn) and the physical properties
of the derivatives both in monomer and in crystalline states.

The electronic and optical properties of pentacene in solution
have been previously calculated using first principles methods
\cite{Tiago-DFT}. The calculated optical spectra of the vapor phase
are found to be in agreement with the measurements performed
on the thin film phase of pentacene. Doi \emph{et al.} \cite{Doi-DFT}
have calculated the electronic band structures for both the single
crystal and thin film polymorphs of pentacene and concluded that the effective mass
of the electrons or holes is larger in the single crystal. A first principles simulation
of the thin film phase of pentacene shows a crucial dependence of the bandwidths
of the HOMO and LUMO and of the band gap on the molecular stacking angles
\cite{Parisse}. The electronic structures of iodine- and rubidium-doped
pentacene molecular crystals have also been investigated by ab-initio calculations
based on the ultrasoft pseudopotential method, predicting a metallic
behavior \cite{Shichibu-DFT}. Recently, the structural and electronic
properties of pentacene multilayers on the Ag(111) surface have been studied,
revealing that pentacene has no electronic contribution at the Fermi level \cite{Mete}. 

\begin{figure*}[t]
\includegraphics[width=0.8\textwidth]{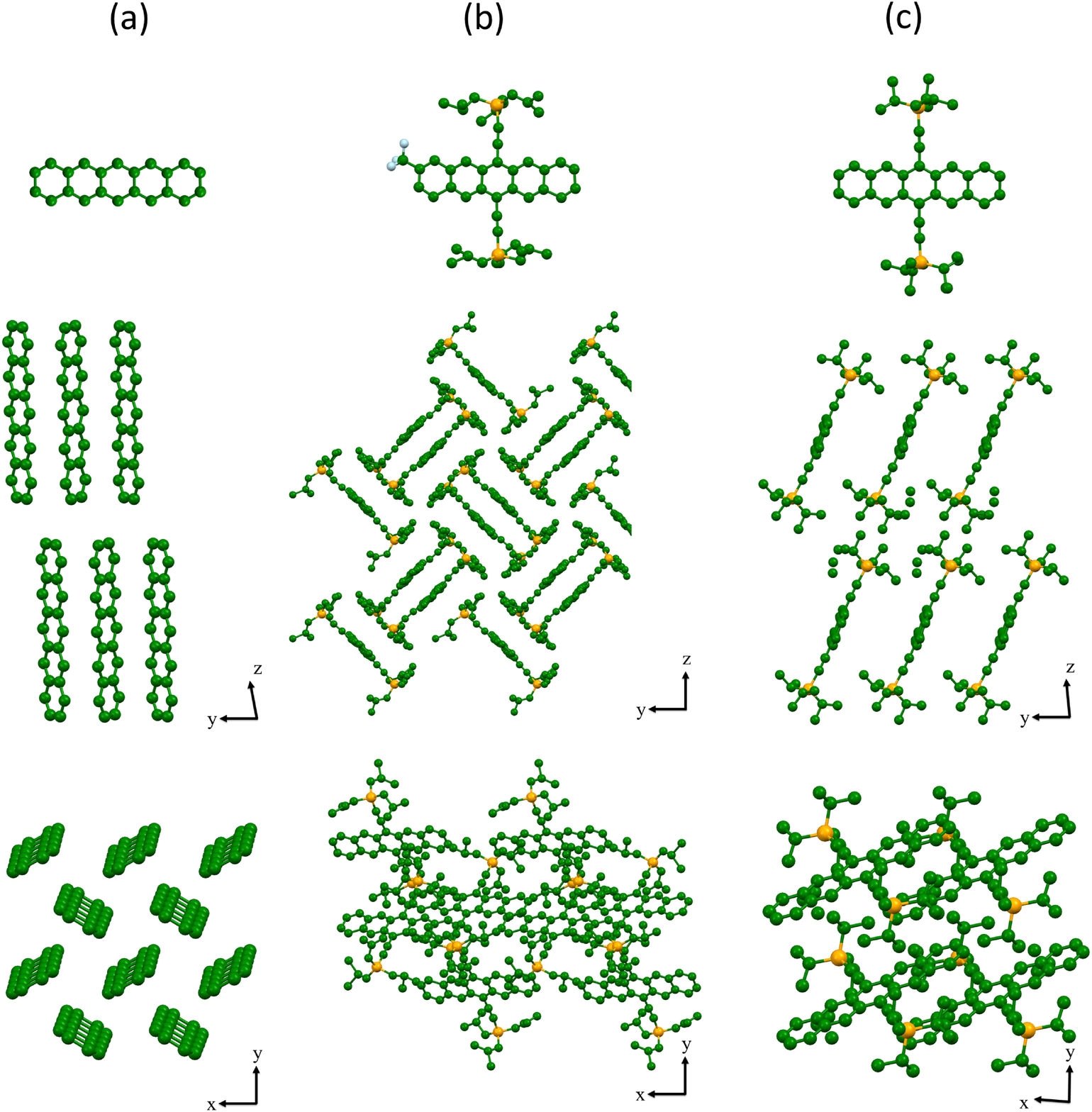}
\caption{Molecular structure (top) and crystal packing (bottom) of (a) pentacene,
(b) TIBS-CF$_3$-Pn, and (c) TIPS-Pn .}
\end{figure*}

Despite several experimental and theoretical investigations, the influence of the
solubilizing chemical substitutions and the resulting changes of the
crystal packing on the electronic and optical properties of pentacene have not been reported. In this work, we
study and compare the theoretical (single molecule and single
crystal) and experimental (dissolved and thin film polycrystal) optical
properties of pentacene, TIPS-Pn, and TIBS-CF$_{3}$-Pn. The results
are analyzed in light of the calculated density of states (DOS) to
see the influence of different alkyl-silyl  groups on the electronic and optical
properties of both monomer and crystal of these materials useful to in electronic and OPV applications.

\section{Experiments and Characterization}

Pentacene, toluene (anhydrous 99.8\%) and 1,3,5-trichlorobenzene (anhydrous 99\%) were purchased from Sigma Aldrich and used without further purification. TIPS-Pn and TIBS-CF$_{3}$-Pn
were synthesized \cite{Chaung-exp-Anthony} and purified by multiple recrystallization 
from acetone (TIPS-Pn) or ethanol (TIBS-CF$_{3}$-Pn). 
Pentacene was dissolved in 1,3,5-trichlobenzene at 100$^\circ$C with a concentration
of 0.5 wt.\% and stirred overnight in the dark. TIPS-Pn and TIBS-CF$_3$-Pn
were dissolved in toluene at room temperature and stirred overnight
in the dark. The solutions were filled in a 1 mm thick quartz cuvette and loaded in
a Cary 5000 (Varian) instrument to aquire UV-vis absorption spectra. 
The measurements were performed over a spectral range from 300
nm to 2000 nm with a 2.0 nm slit width.
Single crystal Si(100) wafers with a thermal oxide
layer of 100 nm thickness were used as substrate for the thin film deposition.
Prior to deposition, the substrates were cleaned in amonium hydroxide (30\% NH$_4$OH),
hydrogen peroxide (30\% H$_2$O$_2$) and Milli Q (1:1:5 ratio) for 15 min at
70$^\circ$C. Thin films of TIPS-Pn and TIBS-CF$_3$-Pn
were spin cast at 1000 rpm for 30 seconds in a N$_2$-filled
glove-box and left to dry in inert atmosphere at room temperature.
 The optical properties of the spin-coated
films were measured using variable angle spectroscopic ellipsometry
(VASE) based on the M-2000XI rotating compensator configuration (J.\
A.\ Woollam Co.\ Inc). VASE spectra ranging from 0.734 eV to 5.895 eV were
recorded at a 18$^\circ$ angle of incidence with 
respect to the substrate normal from 45$^\circ$ to 80$^\circ$ with 2$^\circ$ increment. 
In the paper, we focus on the spectral range from 1 eV to 3 eV. Optical
analysis of VASE data was performed using the EASETM and WVASE32 software packages (J.\
A.\ Woollam Co.\ Inc). Optical modeling was performed assuming a homogeneous
thin film exhibiting uniaxial anisotropy. To describe the dielectric behavior,
a general oscillator approach consisting of Gaussian peaks in the imaginary part
of the dielectric function $\varepsilon_2(E)$ was applied (more detailed information about the fitting procedure and the 
Gaussian parameters can be found in the supporting information). All optical measurements were performed
at room temperature in ambient air.

\section{Simulations}

Our calculations are based on density functional theory, 
using the full-potential linearized augmented plane wave (FP-LAPW)
approach as implemented in the WIEN2k code \cite{wien2k}. This approach
describes the ground state of the present compound with high accuracy
\cite{udo1}. On the other hand, calculation of optical spectra,
in principle, involves excited states. Thus, additional approximations
have to be introduced, which, however, do not compromise the following
line of reasoning \cite{nsingh1}. Exchange and correlation effects
are treated within the local density approximation \cite{LDA}.
In the FP-LAPW method, the unit cell is divided into two parts: non-overlapping
atomic spheres centered at the atomic sites and the interstitial region.
The convergence parameter $R_{mt}\cdot K_{max}$, where $K_{max}$
is the plane wave cut-off and $R_{mt}$ is the smallest of the atomic
sphere radii, controls the size of the basis set. It is set to
$R_{mt}\cdot K_{max}=5$ with $G_{max}=24$. 
A mesh of 48 uniformly distributed \textbf{k}-points in the irreducible wedge of the
Brillouin zone is used for calculating the electronic properties and a dense mesh
of 112 \textbf{k}-points is used to calculate the optical
properties. A total energy convergence of at least $10^{-5}$ Ry is achieved. 
The experimental lattice parameters of pentacene ($a=5.959$ \AA, $b=7.596$ \AA,
$c=15.610$ \AA, $\alpha=81.25^\circ$, $\beta=86.56^\circ$,
and $\gamma=89.90^\circ$), TIPS-Pn ($a=7.565$ \AA, $b=7.750$ \AA, $c=16.835$ \AA,
$\alpha=89.15^\circ$, $\beta=78.42^\circ$, and $\gamma=83.63^\circ$),
and TIBS-CF$_3$-Pn ($a=17.203$ \AA, $b=16.552$ \AA, $c=18.168$ \AA, $\alpha=90^\circ$,
$\beta=113.34^\circ$, and $\gamma=90^\circ$) are used. 

\begin{figure*}[t]
\includegraphics[scale=0.4]{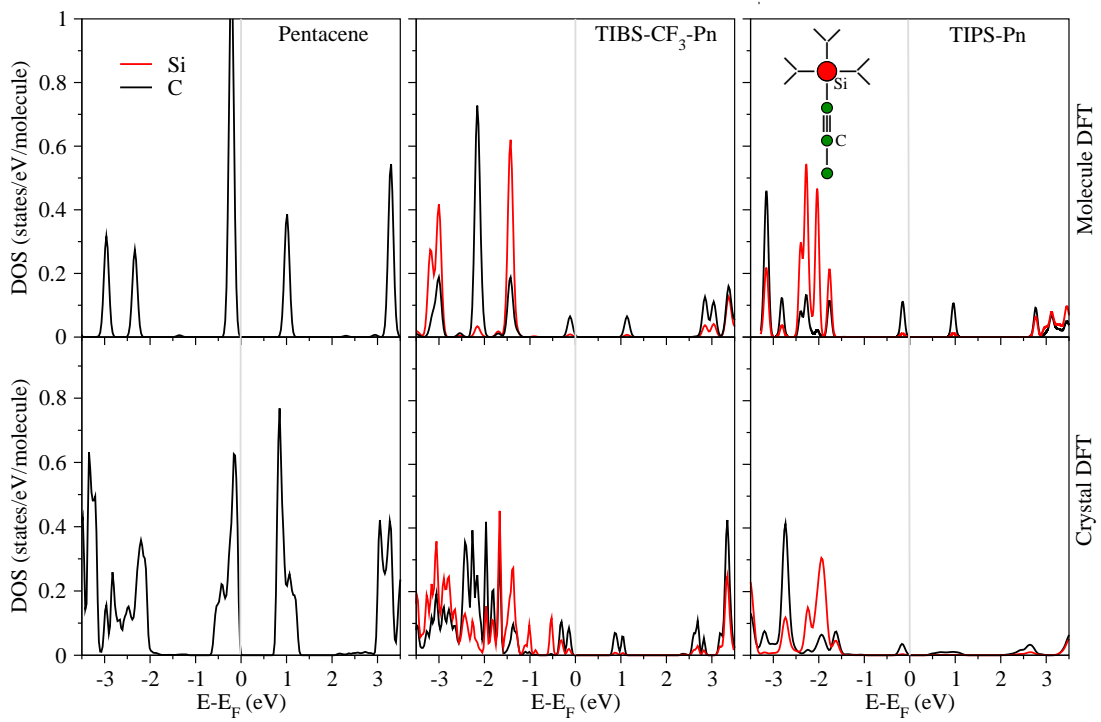}
\caption{Comparison of the pentacene DOS with TIBS-CF$_{3}$-Pn and TIPS-Pn in both molecular
 and crystalline forms.}
\end{figure*}

\section{Results and discussion}

In Fig. 1, we show the molecular and single crystal packing structures of
pentacene, TIPS-Pn, and TIBS-CF$_{3}$-Pn. Pentacene (C$_{22}$H$_{22}$)
and TIPS-Pn (C$_{44}$H$_{54}$Si$_2$) both exhibit triclinic ($P\bar{1}$)
crystal symmetries, while TIBS-CF$_{3}$-Pn (C$_{57}$H$_{65}$F$_3$Si$_2$)
has monoclinic ($P21/c$) symmetry. 

In Fig. 2, we show the calculated projected DOS for pentacene,
TIBS-CF$_{3}$-Pn, and TIPS-Pn for the molecule and crystal
in the energy range $\pm3.5$ eV. The calculated band gaps of pentacene
in the monomer and crystalline phases are found to be 0.84 eV and
0.74 eV, respectively, in agreement with previous
calculations. The low band gap in single crystal pentacene may be
due to the increase in the bandwidth of the HOMO and the LUMO as compared
to the monomer (Fig.\ 2b). The band gaps in TIBS-CF$_{3}$-Pn
and TIPS-Pn are found to be 1 eV and 0.85 eV, respectively, in the monomer
phases, and 0.80 eV and 0.45 eV, respectively, in the single
crystal phases. The band gap of the TIPS-Pn single crystal is the
lowest amongst these molecules owing to the largest LUMO bandwidth amongst the
materials investigated. The calculated band gaps are lower than in experiments, due to the
well known drawback of the local density approximation. The DOS shows
more localized peaks for the pentacene molecule than its derivatives (Fig.\ 2). Due
to the increase of the HOMO and LUMO bandwidths from monomers to crystals,
the bands overlap (below $E_F$) in agreement with previous calculations for pentacene \cite{Doi-DFT}. The HOMO and
LUMO consist mainly of bands belonging to the C and Si atoms
which connect the side group to the pentacene chromophore. The H bands appear 3.5 eV below $E_F$
for pentacene and its derivatives, while the F bands in TIBS-CF$_{3}$-Pn lie
between $-6.0$ eV and $-7.5$ eV (not shown). This means that the electronic
states of the H and F atoms do not contribute to the optical transitions
in the visible energy range and do not participate significantly in
the conversion of sunlight into electricity via absorption. 

\begin{figure*}[t]
\includegraphics[scale=0.4]{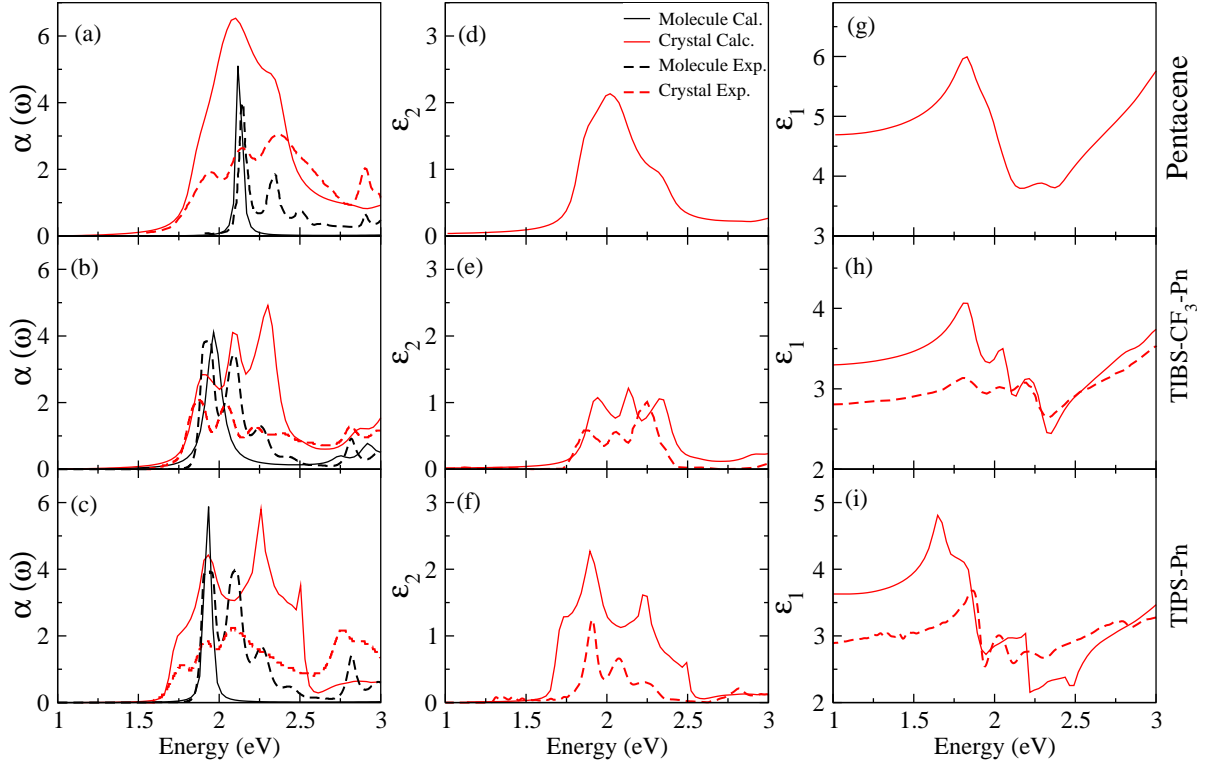}
\caption{Spectra of $\alpha$, $\varepsilon_2$, and $\varepsilon_1$ for pentacene,
TIBS-CF$_{3}$-Pn, and TIPS-Pn in both molecule (monomer) and crystal form
(experiment and simulation).}
\end{figure*}

To establish the effect of the various functional groups on the optical
properties of pentacene and its derivatives, we have calculated their absorption spectra and dielectric
function. The peaks in the optical spectra are
determined by the electric-dipole transitions between the HOMO and
LUMO. Since the local density approximation underestimates the band gap, the calculated spectra 
are shifted by difference between experimental and theoretical band gap to facilitate visual comparison with experimental spectra. In Fig. 3, we present the absorption coefficient $\alpha(E)$ as well as the photon energy-dependent complex dielectric function, $ \widetilde{\varepsilon}=\varepsilon_{1}(E)-i\varepsilon_{2}(E) $. An average of the computed optical spectra along the three coordinate axes is taken and compared with the average of the experimental optical spectra along the axes parallel and perpendicular to the plane of substrate.
The absorption coefficients, $\alpha(E)$, of pentacene and its derivatives are calculated for a single molecule and compared with experimental data of a dilute solution (see Figs. 3(a,b,c)). In the
case of the pentacene solution, the experimental absorption spectrum
shows several peaks at 2.13 eV, 2.30 eV, 2.47 eV, and 2.86
eV. TIBS-CF$_{3}$-Pn and TIPS-Pn monomers exhibit
peaks at 1.94 eV, 2.10 eV, 2.26 eV, 2.45 eV and 2.82 eV. The first absorption
peak in pentacene (2.13 eV) is more intense than the other three peaks, while the first two peaks in TIBS-CF$_{3}$-Pn and TIPS-Pn are more intense than the others. A red shift
of 0.22 eV between the first absorption peak of pentacene and of both TIBS-CF$_{3}$-Pn
and TIPS-Pn, may be due to the effect of the substitutional groups in the latter two derivatives.

The absorption spectrum of pentacene for the thin film exhibits four
absorption peaks at 1.82 eV, 2.13 eV, 2.30 eV, and 2.86 eV which are redshifted
as compare to the monomer. In the
case of TIBS-CF$_{3}$-Pn, the HOMO-LUMO absorption bands are shifted
to lower energies with respect to the monomer spectrum,
having peaks at 1.87 eV, 2.04 eV, 2.22 eV, and 2.81 eV. A similar
phenomenon is observed for TIPS-Pn thin films which show peaks at
1.78 eV, 1.92 eV, 2.08 eV, and 2.76 eV. The shifts of the HOMO-LUMO
absorption band are different in thin films of these materials with respect to the
monomers, namely 0.20 eV for pentacene, 0.07 eV for TIBS-CF$_{3}$-Pn, and
0.16 eV for TIPS-Pn. The different shifts may be attributed to the
different crystalline packing structures. The CF$_{3}$ group does not have
any contribution because the C and F states (see the DOS) are well below the
Fermi level. The calculated and experimental absorption coefficients
also show a redshift between the monomer and the crystal. The calculations for monomers exhibit a single absorption peak, while the experiment shows more than one peak which is consistent with the DOS. The DOS of monomers shows the single sharp LUMO and HOMO bands (allow only one transition peak) while that of for thin film have wider LUMO and HOMO bands, which can have more transitions in absorption spectra. This may be due to the complete isolation of
the molecule in the calculation, which may not be the case in solutions. 

The calculated $\varepsilon_2(E)$ spectra along with their experimental
counterparts for thin film pentacene and its derivatives are presented
in Figs. 3(d,e,f). The experimental $\varepsilon_2(E)$ spectra
of TIBS-CF$_{3}$-Pn and TIPS-Pn show two initial peaks at 1.85 eV and 1.91 eV, respectively, reflecting the optical band gap. The optical band gap is redshifted by 0.06 eV in TIPS-Pn as compared to TIBS-CF$_{3}$-Pn. The third peak is situated at 2.26 eV and 2.19 eV for TIBS-CF$_{3}$-Pn and TIPS-Pn, respectively. 
Another significant difference is the intensity
of the third peak, which dominates in the $\varepsilon_2(E)$ spectrum
of TIBS-CF$_{3}$-Pn while in TIPS-Pn the first peak is most prominent. The $\varepsilon_2(E)$ spectrum changes dramatically by introducing a Si-branch in TIBS-CF$_{3}$-Pn and TIPS-Pn, which is due to the modified crystal
packing. The calculated $\varepsilon_2(E)$ spectra of TIBS-CF$_{3}$-Pn
and TIPS-Pn are in qualitative agreement with our experiments. 

The experimental $\varepsilon_1(E)$ spectra of TIBS-CF$_3$-Pn
and TIPS-Pn thin films show the first transition peaks at energies of
1.80 eV and 1.86 eV, respectively, while second and third peaks position
remain at the same energies in both crystals. This reflects that the alky-silyl length
results in changes of the energy state of the first transition peak in $\varepsilon_1(E)$. The subsequent
peaks at 2.18 eV might be associated with a vibronic energy state
between the Si HOMO and C LUMO. The calculated $\varepsilon_1(E)$ spectra
of crystals and monomers of pentacene, TIBS-CF$_{3}$-Pn, and TIPS-Pn
show similar characteristics. The $\varepsilon_1(E)$ spectra of the TIBS-CF$_{3}$-Pn
and TIPS-Pn crystals demonstrate three peaks similar to the experimental
results. 
Overall, the calculated $\varepsilon_1(E)$ spectra of single crystals of TIBS-CF$_{3}$-Pn and TIPS-Pn are in agreement with our experimental thin film results.

In conclusion, the effects of substitution on the electronic and optical
properties have been discussed based on experiments and theoretical results. 
In the monomer state, the alkyl-silyl substitutions result
in an energy shift of 0.22 eV (experimental) in TIBS-CF$_3$-Pn and TIPS-Pn as compared
to pentacene. In the crystal state, the alkyl-silyl substitution
contributes to different packing structures, which leads to a redshift by 0.09
eV in TIPS-Pn as compared to TIBS-CF$_3$-Pn. The HOMO-LUMO absorption
band in thin films is shifted towards lower energies as compared
to the monomer, by 0.07 eV and 0.16 eV for TIBS-CF$_{3}$-Pn
and TIPS-Pn, respectively. Our first principles calculation of optical spectra have been analyzed
in terms of the calculated DOS. The optical transitions originate primarily from
C and Si bands. A redshift is observed from monomer to crystal for all compounds,
where the extent of redshift depends on the packing structure. Overall, experiment and theory
show a reasonable agreement for the optical spectra.

\newpage

\section{Supporting Information}
The complex dielectric function $\widetilde{\varepsilon}=\varepsilon_{1}-i\varepsilon_{2}$
is related to the complex refraction index $\widetilde{n}=n-ik$ by the following equations:
$\varepsilon_1=n^2-k^2$ and $\varepsilon_2=2nk$. Here, $n$ 
and $k$ are the refractive index and extinction coefficient, respectively.
Kramers-Kronig transformation was used during the model fitting as a constraint: 

\begin{equation}
\varepsilon_1(\omega) = 1 + \frac{2}{\pi} P \int_0^{\infty} \frac{\omega^{'} \epsilon_2 (\omega^{'})}{\omega^{'2} - \omega^2} d\omega^{'}
\end{equation}

\begin{equation}
\varepsilon_2(\omega) = - \frac{2 \omega}{\pi} P \int_0^{\infty} \frac{\epsilon_1 (\omega^{'}) - 1}{\omega^{'2} - \omega^2} d\omega^{'}
\end{equation}

The mean square error was used to quantify the difference between experimental
and model-generated data:

\begin{equation}
MSE = \sqrt{\frac{1}{3n-m}\sum_{i=1}^{n}\left[(N_{E_{i}} - N_{G_{i}})^2 + (C_{E_{i}} - C_{G_{i}})^2 + (S_{E_{i}} - S_{G_{i}})^2 \right]} \times 1000
\end{equation}
where $n$ is the number of wavelengths, $m$ is the number of fit parameters,
and $N=\cos(2\Psi)$, $C=\sin(2\Psi)\cos(\Delta)$, $S=\sin(2\Psi)\sin(\Delta)$.
Where, $\Psi$ and $\Delta$ are the amplitude ratio and phase shift, respectively.

The $MSE$ generated is 12.12 and 15.5 for TIPS-Pn and TIBS-CF$_3$-Pn with all angles
variable from 45$^\circ$ to 80$^\circ$, with 2$^\circ$ increment, respectively.

Gaussian oscillators produce a Gaussian line shape in $\varepsilon_{2}$:
\begin{center}
\begin{equation}
\begin{split}
&\varepsilon_{2}=\sum_{i}^{n}A_{n}\Bigg(\left[\Gamma(\dfrac{E-E_n}{\sigma_n})+\Gamma(\dfrac{E+E_n}{\sigma_n})\right]+ \\ 
& i \cdot \left(\exp\left[-(\dfrac{E-E_n}{\sigma_n})^2\right]+\exp\left[-(\dfrac{E+E_n}{\sigma_n})^2\right]\right)\Bigg)
\end{split}
\end{equation}
\end{center}

where $\sigma_n=B_n/(2\sqrt{\ln(2)})$ and
$n$ is the oscillator number, $A_n=\varepsilon_2(E_n)$ is the amplitude, $E_n$ (eV)
is the center energy and $B_n$ (eV) is the full width at half maximum of the peak. 
The function $\Gamma$ is a convergence series that produces a Kramers-Kronig consistent
line shape for $\varepsilon_1$.

\clearpage

\begin{table*}
\caption{Parameters of the modified Gaussian model obtained by fitting the
imaginary part of dielectric function $\varepsilon_2(E)$ of TIBS-CF$_3$-Pn.} 
\begin{tabular}{lll|lll}
\hline
    \multicolumn{3}{c}{$\varepsilon_{2xx}(E)$=$\varepsilon_{2yy}(E)$} & \multicolumn{3}{c}{$\varepsilon_{2zz}(E)$}\\
    \multicolumn{3}{c}{$\varepsilon_\infty$=1.901$\pm$0.006} & \multicolumn{3}{c}{$\varepsilon_{\infty}$=2.052$\pm$0.015}\\
    \multicolumn{3}{c}{UV pole amplitude=11.361$\pm$0.315} & \multicolumn{3}{c}{UV pole amplitude=4.831$\pm$0.546}\\
    \multicolumn{3}{c}{UV pole energy=6.883$\pm$0.024} & \multicolumn{3}{c}{UV pole energy=6.663$\pm$0.073}\\
\hline
A$_{1}$=0.887$\pm$0.009 &  B$_{1}$=0.125$\pm$0.001 &  E$_{1}$=1.868$\pm$0.001 &  A$_{1}$=0.285$\pm$0.009 &  B$_{1}$=0.159$\pm$0.007 &  E$_{1}$=1.872$\pm$0.003\\
A$_{2}$=0.395$\pm$0.002 &  B$_{2}$=0.115$\pm$0.001 &  E$_{2}$=2.040$\pm$0.000 &  A$_{2}$=0.255$\pm$0.021 &  B$_{2}$=0.127$\pm$0.013 &  E$_{2}$=2.050$\pm$0.008\\
A$_{3}$=0.156$\pm$0.002 &  B$_{3}$=0.137$\pm$0.002 &  E$_{3}$=2.204$\pm$0.001 &  A$_{3}$=0.563$\pm$0.037 &  B$_{3}$=0.155$\pm$0.011 &  E$_{3}$=2.244$\pm$0.006\\
A$_{4}$=0.123$\pm$0.002 &  B$_{4}$=0.206$\pm$0.005 &  E$_{4}$=2.368$\pm$0.004 &  A$_{4}$=3.225$\pm$0.028 &  B$_{4}$=0.670$\pm$0.003 &  E$_{4}$=3.873$\pm$0.008\\
A$_{5}$=0.133$\pm$0.003 &  B$_{5}$=0.510$\pm$0.018 &  E$_{5}$=2.838$\pm$0.006 &  A$_{5}$=2.177$\pm$0.099 &  B$_{5}$=0.125$\pm$0.034 &  E$_{5}$=3.795$\pm$0.002\\
A$_{6}$=0.110$\pm$0.015 &  B$_{6}$=0.119$\pm$0.001 &  E$_{6}$=3.304$\pm$0.006 &  A$_{6}$=0.252$\pm$0.034 &  B$_{6}$=0.337$\pm$0.017 &  E$_{6}$=4.569$\pm$0.007\\
A$_{7}$=0.453$\pm$0.002 &  B$_{7}$=0.119$\pm$0.001 &  E$_{7}$=3.744$\pm$0.000 &  A$_{7}$=0.308$\pm$0.019 &  B$_{7}$=0.657$\pm$0.081 &  E$_{7}$=4.993$\pm$0.018\\
A$_{8}$=0.629$\pm$0.001 &  B$_{8}$=0.989$\pm$0.002 &  E$_{8}$=4.187$\pm$0.002 &  A$_{8}$=0.605$\pm$0.005 &  B$_{8}$=0.975$\pm$0.069 &  E$_{8}$=5.847$\pm$0.014\\
A$_{9}$=0.235$\pm$0.002 &  B$_{9}$=0.814$\pm$0.011 &  E$_{9}$=5.076$\pm$0.006 &              &            &        \\
A$_{10}$=0.082$\pm$0.006 &  B$_{10}$=0.145$\pm$0.013 & E$_{10}$=5.416$\pm$0.006   &              &            &    \\
A$_{11}$=0.617$\pm$0.004 &  B$_{11}$=0.531$\pm$0.006 & E$_{11}$=5.649$\pm$0.003   &              &            &    \\
\hline 
\end{tabular}
\label{table:TIBS} 
\end{table*}

\begin{table*}
\caption{Parameters of the modified Gaussian model obtained by fitting the imaginary part
of dielectric function $\varepsilon_2(E)$ of TIPS-Pn.}
\begin{tabular}{lll|lll} 
\hline
    \multicolumn{3}{c}{$\varepsilon_{2xx}(E)$=$\varepsilon_{2yy}(E)$} & \multicolumn{3}{c}{$\varepsilon_{2zz}(E)$}\\
    \multicolumn{3}{c}{$\varepsilon_{\infty}$=2.012$\pm$0.030} & \multicolumn{3}{c}{$\varepsilon_{\infty}$=2.251$\pm$0.023} \\
    \multicolumn{3}{c}{UV pole amplitude=4.360$\pm$1.999} & \multicolumn{3}{c}{UV pole amplitude=3.889$\pm$0.443} \\
    \multicolumn{3}{c}{UV pole energy=7.139$\pm$0.139} & \multicolumn{3}{c}{UV pole energy=6.190$\pm$0.033} \\
\hline
A$_{1}$=0.830$\pm$0.002 &  B$_{1}$=0.080$\pm$0.002 &  E$_{1}$=1.906$\pm$0.002 &  A$_{1}$=1.087$\pm$0.027 &  B$_{1}$=0.105$\pm$0.003 &  E$_{1}$=1.891$\pm$0.001 \\
A$_{2}$=0.353$\pm$0.004 &  B$_{2}$=0.080$\pm$0.001 &  E$_{2}$=2.069$\pm$0.000 &  A$_{2}$=0.666$\pm$0.028 &  B$_{2}$=0.111$\pm$0.006 &  E$_{2}$=2.053$\pm$0.006 \\
A$_{3}$=0.266$\pm$0.002 &  B$_{3}$=0.497$\pm$0.005 &  E$_{3}$=2.223$\pm$0.002 &  A$_{3}$=1.566$\pm$0.055 &  B$_{3}$=0.154$\pm$0.005 &  E$_{3}$=3.503$\pm$0.002 \\
A$_{4}$=0.508$\pm$0.005 &  B$_{4}$=0.198$\pm$0.003 &  E$_{4}$=4.139$\pm$0.007 &  A$_{4}$=2.709$\pm$0.203 &  B$_{4}$=0.132$\pm$0.011 &  E$_{4}$=3.886$\pm$0.005 \\
A$_{5}$=1.232$\pm$0.314 &  B$_{4}$=0.204$\pm$0.003 &  E$_{5}$=4.236$\pm$0.009 &  A$_{5}$=1.532$\pm$0.141 &  B$_{5}$=0.159$\pm$0.012 &  E$_{5}$=4.256$\pm$0.013 \\
A$_{6}$=2.963$\pm$0.028 &  B$_{6}$=0.467$\pm$0.001 &  E$_{6}$=5.666$\pm$0.006 &  A$_{6}$=1.344$\pm$0.047 &  B$_{6}$=0.234$\pm$0.004 &  E$_{6}$=4.132$\pm$0.016 \\
A$_{7}$=0.396$\pm$0.004 &  B$_{7}$=0.963$\pm$0.022 &  E$_{7}$=3.506$\pm$0.002 &  A$_{7}$=0.467$\pm$0.005 &  B$_{7}$=1.954$\pm$0.045 &  E$_{7}$=5.419$\pm$0.006 \\
A$_{8}$=0.939$\pm$0.009 &  B$_{8}$=0.586$\pm$0.007 &  E$_{8}$=3.868$\pm$0.003 &  A$_{8}$=0.400$\pm$0.045 &  B$_{8}$=0.160$\pm$0.018 &  E$_{8}$=5.737$\pm$0.008 \\
                &                  &           &  A$_{9}$=0.835$\pm$0.020 &  B$_{9}$=0.453$\pm$0.141 & E$_{9}$=4.088$\pm$0.008    \\
\hline
\end{tabular}
\label{table:TIPS}
\end{table*}


\begin{thebibliography}{26}

\bibitem{Zaumseil}Z. Zaumseil and H. Sirringhaus, Chem. Rev. 107, 1296 (2007). 

\bibitem{WANG}H. Wang and D. Yan, NPG Asia Mater. 2, 69 (2010). 

\bibitem{DODABALAPOR}A. Dodabalapor, H. E. Katz, L. Torsi, and R.
C. Haddon, Science 269, 1560 (1995).

\bibitem{Li}X. Li, B. K. C. Kjellander, J. E. Anthony, C. W. M.
Bastiaansen, D. J. Broer, and G. H. Gelinck, Adv. Funct. Mater. 19, 3610 (2009). 

\bibitem{Jurchescu}O. D. Jurchescu, S. Subramanian, R. J. Kline,
S. D. Hudson, J. E. Anthony, T. N. Jackson, and D. J. Gundlach, Chem.
Mater. 20, 6733 (2008). 

\bibitem{Kim}S. H. Kim, M. Jang , H. Yang, J. E. Anthony, and C.
E. Park, Adv. Funct. Mater. 21, 2198 (2011).

\bibitem{Anthony}J. E. Anthony, J. Gierschner, C. A. Landis, S.
R. Parkin, J. B. Sherman, and R. C. Bakus, Chem. Commun. 45, 4746 (2007).

\bibitem{Subramanian}S. Subramanian, S. K. Park, S. R. Parkin, V.
Podzorov, T. N. Jackson, and J. E. Anthony, J. Am. Chem. Soc. 123, 9482 (2001).

\bibitem{Shu}Y. Shu, PhD thesis, University of Kentucky (2011).

\bibitem{Ostroverkhova}O. Ostroverkhova, D. G. Cooke, F. A. Hegmann,
R. R. Tykwinski, S. R. Parkin, and J. E. Anthony, Appl. Phy. Lett.
89, 192113 (2006).

\bibitem{Jackson}S. K. Park, T. N. Jackson, J. E. Anthony, and D.
A. Mourey, Appl. Phys. Lett. 91, 063514 (2007).

\bibitem{Bao-Nature}G. Giri, E. Verploegen, S. C. B. Mannsfeld, S.
A. Evrenk, D. H. Kim, S. Y. Lee, H. A. Becerril, A. A. Guzik, M. F.
Toney, and Z. Bao, Nature 480, 504 (2011).

\bibitem{Meng}H. Meng, M. Bendikov, G. Mitchell, R. Helgeson, F.
Wudl, Z. Bao, T. Siegrist, C. Kloc, and C. H. Chen, Adv. Mater. 15, 1090 (2003). 

\bibitem{Hinderhofer}A. Hinderhofer, U. Heinemeyer, A. Gerlach, S.
Kowarik, R. M. J. Jacobs, Y. Sakamoto, T. Suzuki, and F. Schreiber,
J. Chem. Phys. 127, 194705 (2007). 

\bibitem{lim}Y.-F. Lim, Y. Shu, S. R. Parkin, J. E. Anthony, and
G. G. Malliaras, J. Mater. Chem. 19, 3049 (2009).

\bibitem{Tiago-DFT}M. L. Tiago, J. E. Northrup, and S. G. Louie,
Phys. Rev. B 67, 115212 (2003). 

\bibitem{Doi-DFT}K. Doi, K. Yoshida, H. Nakano, A. Tachibana, T.
Tanabe, Y. Kojima, and K. Okazaki, J. Appl. Phys. 98, 113709 (2005).

\bibitem{Parisse}P. Parisse, L. Ottaviano, B. Delley, and S. Picozzi,
J. Phys.: Condens. Matter 19, 106209 (2007).

\bibitem{Shichibu-DFT}Y. Shichibu and K. Watanabe, Jpn. J. Appl.
Phys. 42, 5472 (2003).

\bibitem{Mete}E. Mete, I. Demiroglu, M. F. Danisman, and S. Ellialtioglu,
J. Phys. Chem. C 114, 2724 (2010).

\bibitem{Chaung-exp-Anthony}Y. S. Chung, N. Shin, J. Kang, Y. Jo,
V. M. Prabhu, S. K. Satija, R. J. Kline, D. M. DeLongchamp, M. F.
Toney, M. A. Loth, B. Purushothaman, J. E. Anthony, and D. Y. Yoon,
J. Am. Chem. Soc., 133, 412 (2011).

\bibitem{wien2k}P. Blaha, K. Schwarz, G. Madsen, D. Kvasicka, and
J. Luitz, WIEN2k, An Augmented Plane Wave + Local Orbitals Program
for Calculating Crystal Properties (TU Vienna, Vienna, 2001).

\bibitem{udo1}U. Schwingenschl\"ogl and C. Schuster, Phys. Rev. Lett. 99, 237206 (2007);
EPL 79, 27003 (2007). 

\bibitem{nsingh1}N. Singh and U. Schwingenschl\"ogl,
Chem. Phys. Lett. 508, 29 (2011).

\bibitem{LDA}J. P. Perdew and A. Zunger, Phys. Rev. B 23, 5048 (1981).

\bibitem{McCullough}R. D. McCullough, Adv. Mater. 10, 93 (1998) 
\end{thebibliography}
\end{document}